\documentstyle[11pt] {article}

\title{Relative information encoded in the degree of entanglement to discriminate bipartite states}
\author{  T. V\'ertesi \\
Institute of Nuclear Research of the Hungarian Academy of Sciences,\\
H-4001 Debrecen, P.O. Box 51, Hungary\\
e-mail:tvertesi@dtp.atomki.hu}

\begin{document}
\maketitle
\newcommand{\beq} {\begin{equation}}
\newcommand{\enq} {\end{equation}}
\newcommand{\ber} {\begin{eqnarray}}
\newcommand{\enr} {\end{eqnarray}}


\begin{abstract}

It has been recently shown (Bartlett et al. 2003) that information encoded
into relative degrees of freedom enables communication without a common
reference frame using entangled bipartite states. In this case the relative
information stored in the two-qubit system is shared between the
polarization degrees of freedom and the degree of entanglement. In the
present article a specific state discrimination problem is envisioned where
the degree of entanglement carries the only relative parameter, so that
certain maximally entangled states are perfectly distinguishable, while
discrimination of product states is impossible.

\end{abstract}

\section{Introduction}

In a composite system, be it classical or quantal, made up of two or more
distinct parts, one can distinguish between global and relative parameters.
Global parameters belong to some properties of the whole system with respect
to an external reference system, while relative parameters describe
relations between the individual parts of the system. On the other,
measuring the relative degrees of freedom of a system exhibits large
differences between the classical and the quantal settings
\cite{Gisin,Bagan}: in the classical world one cannot measure relative
parameters without actually measuring the global parameters as well, while
in the quantum world it is possible to perform measurements revealing merely
relative properties of the system. (Moreover, it has been found
\cite{Bartlett04,Bagan,Lindner} that an optimal measurement of the relative
angle between two 'quantum axes' made out of $N1$ and $N2$ particles can be
performed without the need of an external reference frame.) This also
implies that if one try to encode information in the quantal case in the
relational degrees of freedom, this information will not depend on global
effects acting on the system as a whole. Indeed, this kind of relative
information is useful in quantum computing allowing to design noiseless
quantum codes \cite{Zanardi97,Byrd,Rudolph}, in quantum cryptography
permitting to distribute quantum key in the presence of arbitrary degree of
collective noise \cite{Boileau,Walton}, or in quantum communication if no
shared reference frame (Cartesian or Lorentz) exists between the
communicating parties \cite{Bartlett03,Bourennane,Enk,Bartlett05}.

On the other hand, relative information is also a useful resource in
classical communication over a quantum channel if either the noise on the
quantum channel has some partial correlations
\cite{Macchiavello,Ball,Banaszek} or when the communicating parties do not
share a common reference frame \cite{Bartlett03}. In this latter setting two
qubits are employed to send a single classical bit of information: The
sender transmits either the maximally entangled, antisymmetric singlet state
$|\psi^-\rangle$, or any state from the symmetric subspace
$I-|\psi^-\rangle\langle\psi^-|$. Then the receiver can distinguish the
above two states with certainty by performing a projective measurement on
the respective antisymmetric and symmetric subspaces. Since the state from
the symmetric subspace can be a product state as well, it is suspected that
entanglement does not only play the role of carrying data encoded into
relative degrees of freedom. In fact in a general two-qubit system,
according to Schmidt's theorem \cite{Schmidt,Peres}, there are three
relative parameters \cite{Bartlett04}: The polarization angle between the
two spins in the Schmidt decomposition, the degree of entanglement and the
phase between the two terms in this decomposition.

Thus, it seems an interesting question to raise whether it is possible to
imagine a physical problem where out of the above three relative parameters
only the degree of entanglement carries the whole information content in the
bipartite system. The main purpose of the present article is to find an
example of a state discrimination problem which fulfills this requirement,
so that maximally entangled input states enable perfect state discrimination
while product input states are completely indistinguishable (section 2). We
also give a proof (in section 3) that in the given example the measured
two-qubit states carry relative information encoded indeed only in the
degree of entanglement, and that they can be distinguished with certainty.
Then (in section 4) this problem is examined in its most generality by
calculating the performance of the measurement for non-maximally entangled
and product states as well.

\section{Searching for distinguishable states}

Our problem is related to the discrimination of separated, unentangled
two-qubit correlated states using non-local measurements \cite{Pryde}. In
that case one party prepares two kind of correlated product states of
spin-1/2 particles, $|\psi_1^{sep}\rangle=|{\bf n}\rangle|{\bf -n}\rangle$
and $|\psi_2^{sep}\rangle=|{\bf n}\rangle|{\bf n}\rangle$, where $|{\bf
n}\rangle$ is chosen randomly and uniformly from the whole Bloch sphere. The
task of the other party (i.e the observer) is to identify the above states
with the best efficiency. Since the single qubit states $|{\bf\pm n}\rangle$
are completely unknown for the observer the identification cannot be
accomplished unambiguously. Actually, the information contained in these
correlated product states is supported only by the relative angles (zero or
$\pi$) between the pair of spins. It has been shown \cite{Pryde}, that in
the optimal joint measurement the average probability $\bar{P}$ to
discriminate correctly $|\psi_1^{sep}\rangle$ and $|\psi_2^{sep}\rangle$ is
$\bar{P}=3/4$. On the other hand, if the input states were not restricted to
product states and $|\psi_1\rangle$ would be the antisymmetric state
$|\psi^-\rangle$, while state $|\psi_2\rangle$ would lie in the symmetric
subspace $I-|\psi^-\rangle\langle\psi^-|$, one could perfectly distinguish
between them by performing a projective measurement onto the symmetric and
antisymmetric subspaces \cite{Bartlett03}. Thus the choice of the following
states
\begin{equation}
|\psi_{j}\rangle = \frac{1}{\sqrt2}(|{\bf n}\rangle|{\bf -n}\rangle \pm
|{\bf -n}\rangle|{\bf n}\rangle) \label{psi12nn}
\end{equation}
satisfies the criterion of perfect discrimination, where the plus sign
stands for $j=1$ (the antisymmetric singlet state), while the minus sign
represents $j=2$ (the symmetric states). It is noted, that the product
states $|{\bf n}\rangle|{\bf n}\rangle$ are also elements of the symmetric
subspace. However, it is completely indifferent to the outcome of the
measurement if we choose from the symmetric subspace either the product
state $|{\bf n}\rangle|{\bf n}\rangle$ or the maximally entangled state
$|\psi_2\rangle$ in Eq.~(\ref{psi12nn}). Hence there is no need that both
distinguishable states be maximally entangled, indicating that in this case
entanglement should be only partially responsible for the discrimination of
the states $|\psi_1\rangle$ and $|\psi_2\rangle$ in Eq.~(\ref{psi12nn})
above (although these states are both maximally entangled). This means that
relative information must be stored here partly in the degree of
entanglement and partly in the relative angles between the spin directions
of the pair of qubits.

Our main goal is to transform away the relative polarization degrees of
freedom thereby leaving us merely with the degree of entanglement as the
single relative parameter. In order to accomplish it, we propose the
following new input states
\begin{equation}
|\psi_{j}^{re}\rangle = \frac{1}{\sqrt2}(|{\bf n}^{re}\rangle|{\bf
-m}^{re}\rangle \pm |{\bf -n}^{re}\rangle|{\bf m}^{re}\rangle) \;,
\label{psi12nm}
\end{equation}
where the plus (minus) sign represents $j=1$ ($j=2$), as before in
Eq.~(\ref{psi12nn}), but now $|{\bf n}^{re}\rangle$ and $|{\bf
m}^{re}\rangle$ are completely uncorrelated and are both chosen from a
uniform distribution over the polar great circle, which is a circle lying in
the $x-z$ plane of the Bloch sphere. These states formally originate from
the states defined in Eq.~(\ref{psi12nn}), by switching in the second qubit
the polarization state $|{\bf n}\rangle$ to $|{\bf m}\rangle$ and picking
$|{\bf n}\rangle$ and $|{\bf m}\rangle$ only from the polar line of the
Bloch sphere (superscript {\it re} refers to the latter restriction that the
corresponding states in the $x-z$ plane are confined to the real vector
space). Notice, that taking account of the complete randomness and lack of
correlation between $|{\bf n}^{re}\rangle$ and $|{\bf m}^{re}\rangle$, the
states in Eq.~(\ref{psi12nm}) no longer support the relative parameters
encoded in the polarization degrees of freedom. Therefore, states in
Eq.~(\ref{psi12nm}) due to the lack of knowledge of the observer about the
points $|{\bf n}^{re}\rangle$ and $|{\bf m}^{re}\rangle$ are represented by
the mixed density operator
\begin{equation}
\rho_j=\int{d{\bf n}d{\bf m} |\psi_j^{re}\rangle\langle\psi_j^{re}|}
\label{rhoj}
\end{equation}
for $j=1,2$, where $d{\bf n}$ and $d{\bf m}$ are uniform measures on the
polar line of the Bloch sphere.

The reason for the restriction to a smaller subset of $|{\bf n}\rangle$ and
$|{\bf m}\rangle $ comes from the need to distinguish between $\rho_1$ and
$\rho_2$ with perfect fidelity. Namely, taking the pair of input states in
Eq.~(\ref{psi12nm}) with $j=1$ and $j=2$, they are related by the unitary
transformation $\sigma_z\otimes I$, where $\sigma_z$ is written in the basis
$\{|{\bf n}^{re}\rangle,|{\bf -n}^{re}\rangle\}$. However, $\rho_j$ in
Eq.~(\ref{rhoj}) is invariant under the orthogonal transformation
$|\psi_j^{re}\rangle\rightarrow U_1\otimes U_2|\psi_j^{re}\rangle$, where
$U_1,U_2\in \mathrm{SO(2)}$. Since $\sigma_z$ $\not\in \mathrm{SO(2)}$, this
implies that $\rho_1$ is not guaranteed to be equal to $\rho_2$. On the
other hand, if $|{\bf n}^{re}\rangle$ and $|{\bf m}^{re}\rangle$ in
Eq.~(\ref{psi12nm}) were allowed to sample the whole Bloch sphere (by having
them taken to be $|{\bf n}\rangle$ and $|{\bf m}\rangle$) the invariance of
$\rho_j$ would extend to any unitary operator of the form $U_1\otimes U_2$,
eventually resulting in $\rho_1=\rho_2$. We hope that, on the contrary, the
states defined by Eq.~(\ref{psi12nm}) will be perfectly distinguishable.
This conjecture will be proved and supported with further calculations in
the next sections.

\section{The case of subset of maximally entangled states}

Using Schmidt's theorem the state of any pure two-qubit system, up to an
irrelevant overall phase factor, may be written as
\begin{equation}
|\psi_j\rangle=e^{-i\beta_j/2}\cos\frac{\alpha_j}{2}|{\bf n}_j\rangle|{\bf
m}_j\rangle + e^{i\beta_j/2}\sin\frac{\alpha_j}{2}|{\bf -n}_j\rangle|{\bf
-m}_j\rangle\;, \label{Schmidt}
\end{equation}
where $j=1,2$ denote the two distinct states to be discriminated from each
other, and $(\alpha_j,\beta_j)$  are the Schmidt parameters, left unchanged
under local $\mathrm{SU(2)}$ transformations. Therefore, beside the
spherical angle between $|{\bf n}_j\rangle$ and $|{\bf m}_j\rangle$,
$\alpha_j$ and $\beta_j$ are relative parameters of the system too
\cite{Bartlett04}.

Note that due to the general treatment, from now on $|\psi_j^{re}\rangle$
denotes a general two-qubit state defined by Eq.~(\ref{Schmidt}) with the
restriction that $|{\bf n}_j\rangle$ and $|{\bf m}_j\rangle$ are taken from
the polar great circle, in contrast to the definition by
Eq.~(\ref{psi12nm}).

If we quantify the pure state entanglement by the concurrence $C$, which is
an operationally well defined measure of entanglement on its own right
\cite{Wootters}, we obtain
\begin{equation}
C_j=4 \mathrm{det}(
\mathrm{Tr}_2|\psi_j\rangle\langle\psi_j|)=\sin^2\alpha_j\;,
\label{concurrence}
\end{equation}
where $\mathrm{Tr}_2$ means tracing over the second qubit from the two-qubit
state, and $|\psi_j\rangle$ is an arbitrary state from Eq.~(\ref{Schmidt}).
Thus the angle $\alpha_j$ solely determines the degree of entanglement in an
arbitrary pure two-qubit state, and it can be immediately read off from
Eq.~(\ref{concurrence}) that $\alpha_j=\{0,\pi\}$ correspond to factorizable
states, while maximally entangled states are obtained for
$\alpha_j=\{\pi/2,3\pi/2\}$ (the latter $\alpha_j$'s also correspond to the
entangled states defined by Eq.~(\ref{psi12nm}) in the previous section).

Let us calculate now the density operator $\rho_j$ in Eq.~(\ref{rhoj}) for
$j=1,2$ explicitly by taking the quantum states $|\psi_j^{re}\rangle$ from
Eq.~(\ref{Schmidt}) with the restriction that $|{\bf n}_j\rangle,|{\bf
m}_j\rangle$ represent random points on the polar great circle. Thus,
$\rho_j(\alpha_j,\beta_j)=\int{d{\bf n}d{\bf
m}|\psi_j^{re}\rangle\langle\psi_j^{re}|}$ in this general case also does
not depend on the relative parameter carried by $|{\bf n}^{re}\rangle$ and
$|{\bf m}^{re}\rangle$, and $\rho_j$ becomes a  function of only the pair of
parameters $(\alpha_j,\beta_j)$.

By performing the integration in Eq.~(\ref{rhoj}) for the input state
$|\psi_j^{re}\rangle$, we obtain in the standard computational basis (i.e.
in the basis of eigenstates of $\sigma_z \otimes \sigma_z$) the following
density matrix
\begin{equation}
\rho_j = \frac{1}{4}\left(
\begin{array}{cccc}
1 & 0 & 0 & -q_j \\
0 & 1 & q_j & 0 \\
0 & q_j & 1 & 0 \\
-q_j & 0 & 0 & 1
\end{array}
\right)\;, \label{rhomat}
\end{equation}
where $q_j=\sin\alpha_j \cos\beta_j$ for $j=1,2$ and thus $q_j \in [-1,1]$.
After diagonalization one arrives at
\begin{equation}
\rho_j=\frac{1}{4}(1-q_j)\left[|\phi^+\rangle\langle\phi^+|+|\psi^-\rangle\langle\psi^-|\right]
+\frac{1}{4}(1+q_j)\left[|\phi^-\rangle\langle\phi^-|+|\psi^+\rangle\langle\psi^+|\right]\;,
\label{diag}
\end{equation}
where the four Bell states $|\psi^\pm\rangle=\frac{1}{\sqrt
2}(|+z\rangle|-z\rangle\pm|-z\rangle|+z\rangle)$,
$|\phi^\pm\rangle=\frac{1}{\sqrt
2}(|+z\rangle|+z\rangle\pm|-z\rangle|-z\rangle)$ appear in the density
matrix, and $|+z\rangle$, $|-z\rangle$ represent points on the north and
south poles, respectively. It is noted that $\rho_j$ in Eq.~(\ref{diag}) is
completely determined by the value of $q_j$. Since $\rho_j$ commutes with
any operator of the from $U_1\otimes U_2$, where $U_1,U_2$ are SO(2)
orthogonals (see Eq.~(\ref{rhoj})), this also applies to the particular case
when $q_j=-1$ or $q_j=+1$. Let us denote the two-dimensional subspaces
corresponding to $q_j=-1$ and $q_j=+1$ in Eq.~(\ref{diag}) by
\begin{eqnarray}
E_1 &=& |\phi^+\rangle\langle\phi^+|+|\psi^-\rangle\langle\psi^-| \nonumber \\
E_2 &=&|\phi^-\rangle\langle\phi^-|+|\psi^+\rangle\langle\psi^+| \;,
\label{POVM}
\end{eqnarray}
respectively. Then $\rho_j$ in Eq.~(\ref{diag}) looks as
\begin{equation}
\rho_j=\frac{1}{4}(1-q_j)E_1 + \frac{1}{4}(1+q_j)E_2\;, \label{rhojop}
\end{equation}
and any state $|\psi_j^{re}\rangle$ in the support of subspaces
$\{E_1,E_2\}$ is mapped by Eq.~(\ref{rhoj}) to the completely mixed state
$\frac{1}{2}I_{2}$ over the respective two dimensional subspaces. This
implies, following the argumentation of Bartlett et al.~\cite{Bartlett04},
that in this case the most informative measurement to estimate
$|\psi_j^{re}\rangle$ is simply the projection of these states onto the
subspaces $\{E_1,E_2\}$. Hence, in our problem the POVM elements
\cite{Peres} of the optimal measurement are the projectors $\{E_1,E_2\}$.

Thus, provided that $|\psi_1^{re}\rangle$ is chosen from the $E_1$ subspace
(i.e. $q_1=-1$), while $|\psi_2^{re}\rangle$ is chosen from the $E_2$
subspace ($q_2=+1$), performing a projective measurement on the subspaces
$E_1$ and $E_2$ will exactly discriminate $|\psi_1^{re}\rangle$ from
$|\psi_2^{re}\rangle$. Note, that $q_j=\pm1$ implies $\sin \alpha_j=\pm 1$
(i.e. the concurrence $C_j$ in Eq.~(\ref{concurrence}) is equal to 1), hence
the set of the perfectly distinguishable states is maximally entangled. This
result can also be obtained by observing that all the states which are in
the support of $\{E_j\}$ can be produced from one of the Bell states by
performing local SO(2) rotations, however, rotations acting on the local
Hilbert spaces of each qubit cannot change the degree of entanglement (which
is maximal for the Bell states).

\section{The general situation}
In this section we discuss the case when the two-qubit states
$|\psi_j^{re}\rangle$ for $j=1,2$ (taken from a restricted state space of
the Schmidt decomposition in Eq.~(\ref{Schmidt})) describe non-maximally
entangled states. Let us introduce the following payoff function
\begin{equation}
\bar{P}=p_1 \mathrm{Tr}[E_1\rho_1] + p_2 \mathrm{Tr}[E_2\rho_2] \;,
\end{equation}
which gives the measure of success of our measurement. Here $p_j$ is the
prior probability that $|\psi_j^{re}\rangle \; j=1,2$ is prepared, which can
be taken $p_1=p_2=1/2$, assuming that the observer has no prior knowledge of
$p_j$. Substituting the explicit values of $\rho_j$, $E_j$ from
Eq.~(\ref{rhojop}) and Eq.~(\ref{POVM}) and $p_j$ from above into the payoff
function $\bar{P}$, one obtains
\begin{equation}
\bar{P}=\frac{1}{2} + \frac{\cos \beta_1 \sin \alpha_1 - \cos \beta_2 \sin
\alpha_2}{4} \;.
\end{equation}
Starting from this compact formula a brief elementary calculation shows,
that for a given pair of $(\alpha_1,\alpha_2)$ $\bar{P}$ is maximized for
the states
\begin{equation}
|\psi_j^{re}\rangle = \cos\frac{\alpha_j}{2}|{\bf n}\rangle|{\bf m}\rangle
\pm \sin\frac{\alpha_j}{2}|{\bf -n}\rangle|{\bf -m}\rangle \;,
\end{equation}
where the signs plus/minus stand for $j=1,2$, meaning that the optimal
states $|\psi_j^{re}\rangle$ are confined to the real state space. In this
case the average payoff $\bar{P}$ is the single function of the degree of
entanglement for the two distinct states and looks as follows
\begin{equation}
\bar{P}=\frac{1}{2} + \left|\frac{\sin \alpha_1 + \sin \alpha_2}{4}\right|
\;.
\end{equation}

We arrived at the main result of our specific parameter estimation problem:
For spin-1/2 pairs restricted to the real vector space the average payoff
$\bar{P}$ of the measurement depends only on the degree of entanglement
(characterized by $\sin\alpha_j,\;j=1,2$). Especially, for product states
$\bar{P}=1/2$, meaning that no information can be gained from the
measurement to discriminate the two given product states from each other.
Therefore, we are allowed to say, that in our state estimation problem,
wherein the prepared states are real valued spin-1/2 pairs, only the degree
of entanglement as relative parameter is responsible for acquiring
information about the prepared states.

It has been shown in Ref.~\cite{Vidal} that for states with the same
symmetry properties as for $\rho_j$ in Eq.~(\ref{rhomat}) the separability
criterion takes the form
\begin{equation}
{\mathrm max}\{|2f_j-1|-1,|2g_j-1|-1\}\leq 0\;,
\end{equation}
with $f_j=\frac{1}{2}(1+ q_j)$ and $g_j=\frac{1}{2}(1- q_j)$, where the
above explicit values have been obtained by establishing a complete
correspondence between the system considered in section V.B. of
Ref.~\cite{Vidal} and our state $\rho_j$ in Eq.~(\ref{rhomat}). Since the
above inequality is satisfied for any parameters of $(\alpha_j,\beta_j)$ the
states $\rho_1$ and $\rho_2$ are both separable. However, it is interesting
to mention that in the joint measurement $\{E_1,E_2\}$ one could achieve
perfect discrimination of maximally entangled pure states, despite the fact
that the measured states $\rho_j$ for $j=1,2$ are always unentangled. This
perfect discrimination would not have been possible if the measurement were
only a local one, performed on each qubit separately even with the aid of
classical communication.

\section{Conclusion}

In section 2 of this work we found a special parameter estimation problem
for two-qubit systems where the degree of entanglement supports the single
parameter of the system. In section 3 it has been proven that choosing
quantum states from a special subset of the Bell states, perfect state
discrimination is possible. Then, in section 4 a payoff function has been
defined and with the aid of it we characterized the efficiency of the
measurement in estimating spin-1/2 pairs restricted to the real vector
space.

Let us mention a duality between the problem of the present work and the
problem of Ref.~\cite{Pryde}. In both cases only one single parameter
carries information about the identity of the states to be estimated. In our
problem the parameter is the degree of entanglement, while in
Ref.~\cite{Pryde} this parameter pertains to the angle between the spin
directions in the product state of the pair of particles. In both cases the
parameters are independent of global properties of the systems, thus they
are called relative parameters. However, one difference may appear in the
'performance' of these relative parameters: while in our task an optimal
joint measurement can perfectly discriminate maximally entangled states, on
the other hand product states in the problem of Ref.~\cite{Pryde} are cannot
be distinguished with perfect fidelity.
\\
\\
\noindent {\Large\bf Acknowledgement}
\\
\\
\noindent The author is indebted to Prof.\hspace{-0.1cm} Robert Englman for
carefully reading the manuscript and also would like to thank
Dr.\hspace{-0.2cm} K\'aroly F. P\'al for valuable discussions.


\begin {thebibliography}9

\bibitem{Gisin}
N. Gisin, S. Iblisdir, arXiv: quant-ph/0507118

\bibitem{Bartlett04}
S.D. Bartlett, T. Rudolph, and R.W. Spekkens, Phys. Rev. A {\bf 70} 032321
(2004)

\bibitem{Bagan}
E. Bagan, S. Iblisdir, R. Munoz-Tapia, arXiv: quant-ph/0508187

\bibitem{Lindner}
N.H. Lindner, P.F. Scudo, and D. Bruss, arXiv: quant-ph/0506223

\bibitem{Zanardi97}
P. Zanardi and M. Rasetti, Phys. Rev. Lett. {\bf 79} 3306 (1997)

\bibitem{Byrd}
M.S. Byrd, D.A. Lidar, L.-A. Wu, P. Zanardi, Phys. Rev A {\bf 71} 052301
(2005)

\bibitem{Rudolph}
T. Rudolph and S.S. Virmani, arXiv: quant-ph/0503151

\bibitem{Boileau}
J.-C. Boileau, D. Gottesman, R. Laflamme, D. Poulin, and R.W. Spekkens,
Phys. Rev. Lett. {\bf 92} 017901 (2004)

\bibitem{Walton}
Z.D. Walton, A.F. Abouraddy, A.V. Sergienko, B.E.A. Saleh, and M.C. Teich,
Phys. Rev. Lett. {\bf 91} 087901 (2003)

\bibitem{Bartlett03}
S.D. Bartlett, T. Rudolph, and R.W. Spekkens, Phys. Rev. Lett. {\bf 91}
027901 (2003)

\bibitem{Bourennane}
M. Bourennane, M. Eibl, S. Gaertner, C. Kurtsiefer, A. Cabello, and H.
Weinfurter, Phys. Rev. Lett. {\bf 92} 107901 (2004)

\bibitem{Enk}
S.J. van Enk, arXiv: quant-ph/0602079

\bibitem{Bartlett05}
S.D. Bartlett, D. R. Terno, Phys. Rev. A {\bf 71} 012302 (2005)

\bibitem{Macchiavello}
C. Macchiavello and G.M. Palma, Phys. Rev. A {\bf 65} 050301 (2002)

\bibitem{Ball}
J. Ball, A. Dragan, and K. Banaszek, Phys. Rev. A {\bf 69} 042324 (2004)

\bibitem{Banaszek}
K. Banaszek, A. Dragan, W. Wasilewski, and C. Radzewicz, Phys. Rev. Lett.
{\bf 92} 257901 (2004)

\bibitem{Schmidt}
E. Schmidt, Ann. Math. {\bf 63} 433 (1907)

\bibitem{Peres}
A. Peres, {\it Quantum Theory: Methods and Concepts} (Kluwer Academic,
Dordrecht, 1995), pp. 123-126

\bibitem{Pryde}
G.J. Pryde, J.L. O'Brien, A.G. White, and S.D. Bartlett, Phys. Rev. Lett.
{\bf 94} 220406 (2005)

\bibitem{Wootters}
W.K. Wootters, Phys. Rev. Lett. {\bf 80} 2245 (1998)

\bibitem{Vidal}
G. Vidal, R.F. Werner, Phys. Rev. A {\bf 65} 032314 (2002)




\end {thebibliography}
\end{document}